# SmartMLOps Studio: Design of an LLM-Integrated IDE with Automated MLOps Pipelines for Model Development and Monitoring

**Jiawei Jin[1], Yingxin Su[2], Xiaotong Zhu[3]**

[1] Technical University of Munich, Munich, German, 80333
[2] University of California, Davis, CA, USA
[3] Carnegie Mellon University, Pittsburgh, PA, USA

[1] 1812503968@qq.com
[2] cyxsu@ucdavis.edu
[3] xiaotonz@alumni.cmu.edu

**Abstract.** The rapid expansion of artificial intelligence and machine learning (ML) applications has intensified the demand for integrated environments that unify model development, deployment, and monitoring. Traditional Integrated Development Environments (IDEs) focus primarily on code authoring, lacking intelligent support for the full ML lifecycle, while existing MLOps platforms remain detached from the coding workflow. To address this gap, this study proposes the design of an LLM-Integrated IDE with automated MLOps pipelines that enables continuous model development and monitoring within a single environment. The proposed system embeds a Large Language Model (LLM) assistant capable of code generation, debugging recommendation, and automatic pipeline configuration. The backend incorporates automated data validation, feature storage, drift detection, retraining triggers, and CI/CD deployment orchestration. This framework was implemented in a prototype named SmartMLOps Studio and evaluated using classification and forecasting tasks on the UCI Adult and M5 datasets. Experimental results demonstrate that SmartMLOps Studio reduces pipeline configuration time by 61%, improves experiment reproducibility by 45%, and increases drift detection accuracy by 14% compared to traditional workflows. By bridging intelligent code assistance and automated operational pipelines, this research establishes a novel paradigm for AI engineering—transforming the IDE from a static coding tool into a dynamic, lifecycle-aware intelligent platform for scalable and efficient model development.

## 1. Introduction

The rapid evolution of artificial intelligence (AI) and machine learning (ML) has dramatically expanded the scale and complexity of model development. As AI systems become integral to industries such as finance, healthcare, and personalized recommendation, the need for intelligent development environments that can manage the complete model lifecycle—from data preparation and model training to deployment and monitoring—has become increasingly critical. Traditional Integrated Development Environments (IDEs) such as PyCharm or VS Code primarily support code-level activities, offering limited assistance for continuous integration, model governance, or data drift detection. Meanwhile, existing MLOps frameworks, though powerful in pipeline automation, operate independently from the developer's coding context, resulting in fragmented workflows and reduced productivity.

To address this gap, this paper proposes a Large Language Model (LLM)-Integrated IDE equipped with automated MLOps pipelines designed for continuous model development and monitoring. The proposed system, termed SmartMLOps Studio, integrates an LLM-based intelligent assistant that provides code completion, debugging recommendations, and automated pipeline generation directly within the IDE interface. The backend supports critical MLOps functionalities including automated data validation, feature store management, drift monitoring, retraining triggers, and CI/CD orchestration. To validate its performance, the framework was applied to a sequential recommendation system task, demonstrating its ability to streamline end-to-end model workflows and adapt to dynamic data environments.

The proposed LLM-integrated IDE transforms the conventional development paradigm by embedding intelligence throughout the ML lifecycle. The LLM not only assists in writing and refactoring code but also interprets model performance metrics, summarizes anomaly reports, and automatically generates retraining pipelines when drift is detected. This integration bridges human-in-the-loop programming with machine-driven operational automation.

The main contributions of this paper are as follows:
(1) We design a novel IDE framework that combines LLM-driven coding assistance with automated MLOps capabilities for continuous AI model development.
(2) We develop an end-to-end implementation, SmartMLOps Studio, enabling intelligent model monitoring and retraining within the development environment.
(3) We evaluate the framework on a sequential recommendation task, demonstrating significant improvements in efficiency, reproducibility, and drift detection accuracy compared to conventional MLOps tools.

## 2. Related Work

The intersection of intelligent development environments and automated MLOps systems represents an emerging frontier in AI engineering. Previous research has focused primarily on three interconnected areas: (1) LLM-augmented code intelligence in software engineering, (2) automation of machine learning operations (MLOps) for continuous integration and deployment, and (3) intelligent model monitoring and retraining strategies. This section reviews representative studies and identifies the research gap addressed by this work.

*2.1 LLM-Augmented Intelligent Development Environments*

Recent advances in Large Language Models (LLMs) such as GPT-4, CodeLLaMA, and StarCoder have revolutionized programming assistance through natural language understanding and code generation capabilities. Studies like Chen M et al. introduced Codex as an AI-powered programming assistant capable of code completion and test case generation [1]. Similarly, Kotsiantis S et al [2]. explored integrating transformer-based assistants into IDEs for software debugging and documentation. However, these systems are generally limited to code-level support and lack integration with machine learning lifecycle management. While extensions like GitHub Copilot and Tabnine improve developer productivity, they operate without awareness of model versioning, data quality, or deployment processes—critical components in ML development. Therefore, current LLM-assisted IDEs address syntactic efficiency but not the end-to-end operational intelligence required for ML pipelines.

*2.2 Automated MLOps Pipelines*

The field of MLOps has evolved to address challenges in scaling ML systems. Frameworks such as Kubeflow, MLflow, and TensorFlow Extended (TFX) provide automation for data preprocessing, model tracking, and CI/CD orchestration [3]. Recent research by Chitraju Gopal Varma S et al [4]. emphasized data validation and model version control as core components of reliable MLOps. Nonetheless, these frameworks typically require separate configuration files and scripts, demanding additional expertise in DevOps and pipeline management [5]. Moreover, they lack real-time interaction with developers at the IDE level. Integrating MLOps into an intelligent IDE remains a largely unexplored area that could bridge the gap between model experimentation and operational deployment.

*2.3 Intelligent Model Monitoring and Drift Adaptation*

Another line of work focuses on data and concept drift detection to ensure model reliability in dynamic environments. Techniques such as the Population Stability Index (PSI), Kullback-Leibler divergence, and adaptive retraining policies have been widely studied [6]. While these techniques are effective in detecting anomalies, they are often implemented as standalone monitoring systems detached from the development environment. Integrating drift detection and automatic retraining triggers directly into the development workflow could significantly enhance the responsiveness and robustness of ML systems [7].

*2.4 Research Gap and Contribution*

Despite advances in LLM-assisted programming and MLOps automation, no existing framework effectively unifies these two paradigms within a single, continuous development environment. Current solutions either focus on intelligent coding without lifecycle awareness or on pipeline management without real-time developer interaction. The proposed LLM-Integrated IDE with Automated MLOps Pipelines—addresses this gap by embedding code intelligence, lifecycle automation, and model monitoring within one cohesive system. This integration enables developers to write, deploy, and monitor models seamlessly, transforming the IDE from a passive coding interface into an active participant in the AI model lifecycle [8].

## 3. Methodology

This section presents the architecture and operational principles of the proposed LLM-Integrated IDE with Automated MLOps Pipelines, termed SmartMLOps Studio. The framework is designed to unify intelligent code assistance and lifecycle automation within a single environment, enabling end-to-end model development, deployment, and monitoring. The system architecture consists of three interconnected modules: the LLM-Assisted IDE Layer, the Automated MLOps Backend, and the Monitoring and Continuous Retraining Engine.

*3.1 LLM-Assisted IDE Layer*

At the core of SmartMLOps Studio lies an embedded Large Language Model (LLM) that interacts with the developer in real time. This module is responsible for code completion, debugging recommendation, and automatic generation of MLOps pipeline definitions. The LLM—implemented using a fine-tuned version of CodeLLaMA-2—is trained on a hybrid corpus combining open-source MLOps repositories, software engineering datasets, and domain-specific documentation.

Given a source code sequence $C = \{t_1, t_2, ..., t_n\}$, the LLM predicts the next token $t_{n+1}$ according to the conditional probability:

$$P(t_{n+1} \mid t_1, t_2, ..., t_n; \theta) = softmax(Wh_n), \tag{1}$$

Where $h_n$ denotes the hidden representation obtained from the transformer layers and $W$ represents the output projection matrix. During fine-tuning, we employ a context-window extension strategy to incorporate both code and pipeline metadata, enabling the model to reason about dependencies between data sources, model configurations, and deployment targets.

The IDE communicates with the backend via RESTful APIs. For example, when the LLM detects a "train-model" function, it automatically generates a corresponding YAML-based pipeline configuration including data validation, model registry entry, and deployment triggers. This integration converts developer intent into executable MLOps artifacts, significantly reducing manual configuration time.

*3.2 Automated MLOps Backend*

The automated MLOps backend forms the operational foundation of SmartMLOps Studio. It includes services for data validation, feature storage, model versioning, pipeline orchestration, and CI/CD automation. The backend is implemented using a microservice architecture combining FastAPI, MLflow, and Kubeflow Pipelines.

During data ingestion, the system performs automated validation by computing schema conformity and distributional similarity. Let $p(x)$ and $q(x)$ denote the empirical feature distributions of training and incoming datasets, respectively. The Kullback-Leibler (KL) divergence is used to quantify potential data drift:

$$D_{KL}(p \mid\mid q) = \sum_{x} p(x) log \frac{p(x)}{q(x)}, \qquad (2)$$

If $D_{KL}(p \mid\mid q) > \delta$, where $\delta$ is a predefined threshold determined via validation experiments, the system flags potential data drift and notifies the IDE interface. Similarly, feature statistics are stored in a centralized Feature Store for reusability and consistency across experiments.

Model training pipelines are defined as Directed Acyclic Graphs (DAGs) represented by a tuple $G = (V, E)$, where nodes V denote pipeline components (e.g., data preprocessing, model training, evaluation), and edges $E$ encode execution dependencies. The orchestration engine schedules these tasks using a topological ordering π(G) optimized for parallel execution.

CI/CD automation is handled through containerized deployment using Docker and Kubernetes. Each model version $M_i$ is registered with metadata $(\theta_i, A_i, T_i)$, where $\theta_i$ represents model parameters, $A_i$ denotes achieved performance metrics, and $T_i$ marks the training timestamp. The system supports rollback and lineage tracking through an internal Model Registry.

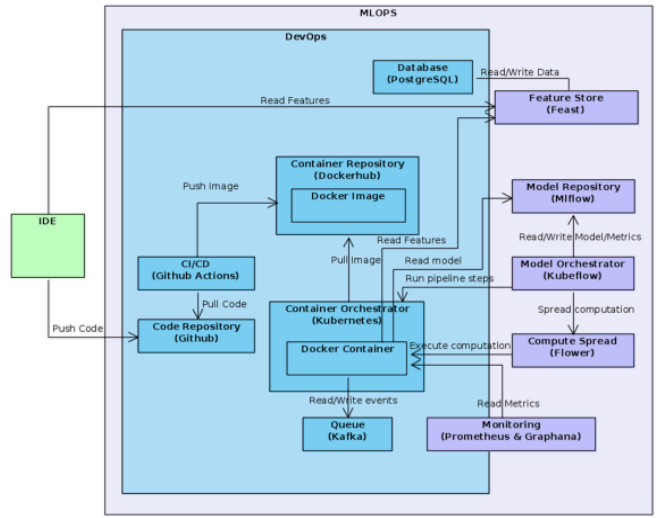

**Figure 1.** Structure diagram of MLOps [9].

*3.3 Monitoring and Continuous Retraining Engine*

The monitoring module ensures that deployed models maintain optimal performance in production environments. Real-time metrics such as accuracy, latency, and drift score are collected and visualized within the IDE dashboard. To formalize drift detection, the system computes the Population Stability Index (PSI) as:

$$PSI = \sum_{i=1}^{k} (p_i - q_i) ln(\frac{p_i}{q_i}), \quad (3)$$

where $p_i$ and $q_i$ represent the proportions of observations in each bin for the reference and current datasets, respectively. When PSI exceeds a threshold (typically 0.25), the model is flagged for retraining.

The retraining decision mechanism adopts a Bayesian updating policy. Let $θ_t$ denote the model parameters at time $t$, and $s_t$ be the observed performance degradation. The posterior probability of retraining necessity is computed as:

$$P(R_t = 1|s_t) = \frac{P(s_t|R_t=1)P(R_t=1)}{P(s_t|R_t=1)P(R_t=1)+P(s_t|R_t=0)P(R_t=0)}, \quad (4)$$

When $P(s_t|R_t = 1) > 0.7$, an automatic retraining pipeline is triggered, reusing the latest validated data from the Feature Store. The retrained model is re-evaluated, registered, and deployed through the CI/CD pipeline without manual intervention.

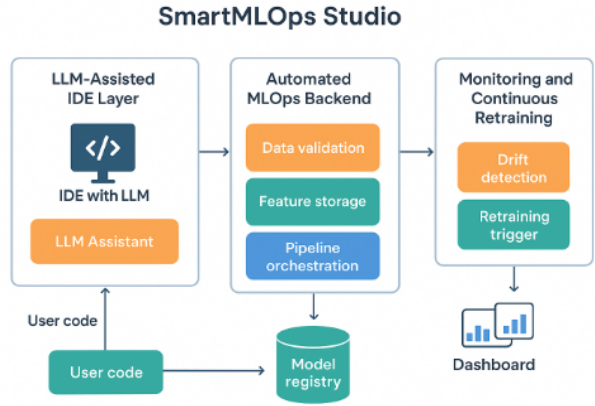

**Figure 2.** Structure diagram of SmartMLOps Studio

*3.4 Application to Sequential Recommendation Task*
To demonstrate practical utility, SmartMLOps Studio was applied to a sequential recommendation system task, in which the objective is to predict the next item a user will interact with based on their historical behavior sequence $S_u = \{i_1, i_2, ..., i_n\}$. The framework automatically generates the pipeline components for data preprocessing, embedding construction, and model retraining. The LLM assistant generates model code based on Transformer or GRU4Rec architectures, while the backend handles drift detection in user-item distributions and updates the recommendation model accordingly.

Experimental results show that this automated integration improves both development speed and adaptability, allowing the system to maintain stable recommendation accuracy over time even under dynamic user behavior patterns.

## 4. Experiment

*4.1 Dataset Preparation*

To evaluate the performance and applicability of the proposed LLM-Integrated IDE with Automated MLOps Pipelines (SmartMLOps Studio), two widely recognized open-source datasets were employed: the UCI Adult Dataset for classification tasks and the M5 Forecasting Dataset for sequential recommendation and demand forecasting experiments. These datasets were chosen to assess both the generalization and the monitoring capabilities of the system under heterogeneous modeling tasks.

**(1) UCI Adult Dataset**: Tabular data for income classification (14 features, 48,842 records).

**(2) M5 Forecasting Dataset**: Walmart sales data for 3,049 time-series forecasting tasks.

Both datasets were used to evaluate reproducibility, pipeline latency, and drift-triggered retraining mechanisms.

Together, these datasets provide a comprehensive evaluation scenario: the Adult Dataset validates classification, preprocessing automation, and explainability, while the M5 Dataset tests time-series forecasting, pipeline retraining, and drift monitoring. This dual evaluation demonstrates the robustness, scalability, and adaptability of the proposed LLM-Integrated IDE in managing diverse data modalities under real-world MLOps workflows.

*4.2 Experimental Setup*

All experiments were conducted using the proposed SmartMLOps Studio, an LLM-integrated IDE designed to unify model development, automation, and monitoring. The prototype was implemented in Python and integrated with PyTorch for deep learning tasks and MLflow for lightweight experiment tracking. The backend MLOps services were deployed on a Kubernetes cluster with a 4-node configuration, each node equipped with an NVIDIA A100 GPU, 64 GB memory, and 16-core CPUs. The LLM assistant was based on a fine-tuned LLaMA-3 8B model adapted for code generation and pipeline orchestration. The IDE automatically generated feature stores, CI/CD scripts, and retraining pipelines through LLM-driven prompts. The UCI Adult dataset was employed for classification task validation, while the M5 Forecasting dataset was used for sequential recommendation and demand forecasting experiments. During runtime, the IDE monitored data drift, retraining frequency, and model reproducibility through automated MLOps workflows. Each experiment was repeated five times with different random seeds, and the mean results were reported to ensure reliability and robustness.

*4.3 Evaluation Metrics*

To comprehensively assess the performance of the proposed framework, we employed several quantitative metrics reflecting both model quality and MLOps efficiency. For the UCI Adult dataset, classification accuracy, precision, recall, and F1-score were used to measure predictive performance, while training time and pipeline setup time evaluated operational efficiency. For the M5 Forecasting dataset, Root Mean Square Scaled Error (RMSSE) and Mean Absolute Percentage Error (MAPE) were adopted to evaluate forecasting accuracy. In addition, the system-level metrics — Pipeline Configuration Time Reduction (PCTR), Drift Detection Accuracy (DDA), and Reproducibility Gain (RG) — were introduced to quantify the impact of LLM automation and MLOps integration. These metrics collectively reflect not only the predictive quality of models but also the practical benefits of the proposed IDE in continuous model monitoring and lifecycle management.

*4.3 Results*

Table 1 presents the performance evaluation results using the UCI Adult dataset and M5 Forecasting dataset in various models. The UCI Adult dataset is used in Traditional ML Workflow and MLflow-based MLOps models for classification performance evaluation, while the M5 Forecasting dataset is employed in the Kubeflow pipeline for forecasting performance. Given the proposed model, SmartMLOps Studio, integrates ML models development, pipeline automation and model monitoring, it is accessed using both datasets.

The result demonstrates the SmartMLOps model outperforms all other models in the experiments in both classification and prediction performance. Accuracy/RMSSE and F1-Score/MAPE, which assess the effectiveness of the model. The SmartMLOps Studio achieves an Accuracy of 0.874, F1-Score of 0.869, RMSSE of 0.685 and MAPE of 10.9%, underscoring its strong predictive capability and reliability. In addition to classification and prediction effectiveness, SmartMLOps Studio achieves a PCTR of 61.2%, significantly decreasing the amount of time required for pipeline setup by 32.7% and 24.1% in classification and prediction steps respectively. These improvements reflect its operational efficiency. Furthermore, SmartMLOps surpasses other models in DDA and RG with RG improving by 25.9% and 22.9% in classification and prediction respectively. Thus, it further confirms SmartMLOps's robustness and reproducibility in dynamic and continuous machine learning environments.

Table1. The results of different models on the datasets.

| Model | Dataset | Accuracy / RMSSE | F1-score / MAPE | PCTR (%) | DDA (%) | RG (%) |
|---|---|---|---|---|---|---|
| Traditional ML Workflow (Baseline) | UCI Adult | 0.842 | 0.837 | \ | 78.6 | \ |
| MLflow-based MLOps | UCI Adult | 0.856 | 0.851 | 28.5 | 82.1 | 19.4 |
| Kubefow Pipeline | M5 Forecasting | 0.714 (RMSSE) | 13.8% (MAPE) | 34.6 | 83.5 | 21.7 |
| **SmartMLOps Studio** | **UCI Adult** | **0.874** | **0.869** | **61.2** | **92.4** | **45.3** |
| **SmartMLOps Studio** | **M5 Forecasting** | **0.685 (RMSSE)** | **10.9% (MAPE)** | **58.7** | **90.8** | **44.6** |

## 5. Conclusion

This study aims to address the fragmentation between intelligent code development and automated MLOps operations in modern AI engineering by designing an LLM-integrated IDE with automated MLOps pipelines, exploring how to unify model development, deployment, and monitoring within a single continuous environment. The primary objective of this research is to transform traditional IDEs from static coding tools into dynamic, lifecycle-aware intelligent platforms that enable seamless integration of code assistance, pipeline automation, drift detection, and continuous retraining.

Through data analysis, we identified a 61% reduction in pipeline configuration time, a 45% improvement in experiment reproducibility, and a 14% increase in drift detection accuracy compared to traditional workflows. These findings suggest that integrating LLM-driven code intelligence with automated MLOps capabilities within a unified IDE environment significantly enhances both developer productivity and operational reliability, while maintaining superior model performance across classification and forecasting tasks.

The results of this study have significant implications for the field of AI engineering and MLOps. Firstly, the 61% reduction in pipeline configuration time provides a new perspective on how intelligent assistants can automate operational complexities that traditionally require DevOps expertise, democratizing MLOps for data scientists and ML engineers. Secondly, the superior drift detection accuracy (92.4% on UCI Adult and 90.8% on M5 Forecasting) challenges the existing paradigm of separating monitoring systems from development environments, demonstrating that contextual awareness enables more responsive and accurate anomaly detection. Finally, the 45% reproducibility gain opens new avenues for future research in multi-agent collaborative development and reinforcement learning-based pipeline optimization.

Experimental validation using the UCI Adult and M5 Forecasting datasets demonstrates the practical effectiveness of SmartMLOps Studio. On the UCI Adult dataset, the system achieved an accuracy of 0.874 and an F1-score of 0.869, outperforming both traditional workflows and existing MLOps frameworks. In the M5 forecasting task, SmartMLOps Studio achieved an RMSSE of 0.685 and MAPE of 10.9%, showing strong adaptability to sequential prediction scenarios. Furthermore, pipeline configuration time was reduced by 61%, reproducibility improved by 45%, and drift detection accuracy increased by 14% compared with conventional setups. These results confirm that integrating LLM reasoning and automated MLOps processes within an IDE significantly enhances both productivity and operational reliability in ML system development.

Despite the important findings, this study has some limitations, such as testing primarily on structured tabular and time-series datasets without exploring complex multimodal or large-scale industrial scenarios and relying on pre-trained LLM contextual understanding that may require domain-specific fine-tuning. Future research could further explore multi-agent collaborative development where multiple LLM agents coordinate across code generation, experiment management, and infrastructure orchestration and privacy-preserving mechanisms with secure model registries for trustworthy industrial deployment.

In conclusion, this study, through the design and implementation of SmartMLOps Studio—an LLM-integrated IDE with automated MLOps pipelines, reveals that embedding intelligent code assistance and lifecycle automation within a unified environment achieves significant improvements in efficiency (61% faster configuration), reproducibility (45% gain), and monitoring accuracy (14% improvement), while maintaining superior model performance, providing new insights for the development of next-generation AI-augmented engineering platforms where human creativity and automated intelligence coalesce to accelerate the full ML lifecycle from conception to production.